# A novel approach for classifying Monoamine Neurotransmitters by applying Machine Learning on UV plasmonic-engineered Auto Fluorescence Time Decay Series (AFTDS)

Mohammad Mohammadi[1], Sima Najafzadehkhoei[2], George Vega Yon[2,3], Yunshan Wang[1*]

[1] Department of Chemical Engineering, University of Utah, Salt Lake City, UT, USA

[2] Department of Population Health Sciences, University of Utah, Salt Lake City, UT, USA

[3] Department of Internal Medicine, University of Utah, Salt Lake City, UT, USA

**Abstract**

This study introduces a hybrid approach integrating advanced plasmonic nanomaterials and machine learning (ML) for high-precision biomolecule detection. We leverage aluminum concave nanocubes (AlCNCs) as an innovative plasmonic substrate to enhance the native fluorescence of neurotransmitters, including dopamine (DA), norepinephrine (NE), and 3,4-Dihydroxyphenylacetic acid (DOPAC). AlCNCs amplify weak fluorescence signals, enabling probe-free, label-free detection and differentiation of these molecules with great sensitivity and specificity. To further improve classification accuracy, we employ ML algorithms, with Long Short-Term Memory (LSTM) networks playing a central role in analyzing time-dependent fluorescence data. Comparative evaluations with k-Nearest Neighbors (KNN) and Random Forest (RF) demonstrate the superior performance of LSTM in distinguishing neurotransmitters. The results reveal that AlCNC substrates provide up to a 12-fold enhancement in fluorescence intensity for DA, 9-fold for NE, and 7-fold for DOPAC compared to silicon substrates. At the same time, ML algorithms achieve classification accuracy exceeding 89%. This interdisciplinary methodology bridges the gap between nanotechnology and ML, showcasing the synergistic potential of AlCNC-enhanced native fluorescence and ML in biosensing. The framework paves the way for probe-free, label-free biomolecule profiling, offering transformative implications for biomedical diagnostics and neuroscience research.

**Keywords:** Aluminum, Concave Nano Cubes, Neurotransmitter, Plasmonic, Machine Learning

* Corresponding Author: yunshan.wang@utah.edu

**Introduction**

The precise quantification and identification of monoamine neurotransmitters (MANTs) play a pivotal role in understanding neurological processes and in the early diagnosis of neurodegenerative diseases[1-4]. Conventional analytical methods such as chromatography and mass spectrometry [5] require complex sample preparation and expensive reagents, unsuitable for frequent assessments of MANT levels [6, 7]. Electrochemical approaches such as the Fast-Scan Cyclic Voltammetry (FSCV) method [8] are cost-effective alternatives; however, they cannot distinguish MANTs with similar oxidation potentials [9]. Genetically encoded fluorescence probes [10, 11] are able to detect MANTs with excellent sensitivity and specificity; however, they require the use of transgenic animals. Antibody or aptamer-based assays [12, 13] have demonstrated real-time sensing of MANTs. However, the long-term stability of these probes in biological fluids remains an issue [14-17].

MANTs have an aromatic ring structure that emits autofluorescence (AF) when excited by ultraviolet (UV) light. The AF absorption cross-sections are orders of magnitude higher than that of Raman or Infrared absorption. Therefore, AF spectroscopy is a promising technique for sensitive, label-free, and probe-free quantification and identification of MANTs. However, the classification of similar MANTs based on their AF profile is challenging due to their overlapping spectrum [18]. We have shown in our prior work that the AF of MANTs drop cast on a solid substrate (e.g., a silicon wafer or a plasmonic nano hole array, etc.) decays exponentially over time when continuously exposed to UV light [19-22]. The decay rate constants were found to be distinct among similarly structured MANTs, and their differences were enlarged by a UV plasmonic nano hole-array [20]. Our prior work focused solely on the decay rate constants and has not realized the full potential of using UV plasmonic-engineered Auto Fluorescence Time Decay Series (AFTDS) in classifying MANTs. In this paper, we demonstrated excellent classification accuracy by combining artificial intelligence with the AFTDS of MANTs deposited on aluminum plasmonic concave nanocubes (AlCNC). The assembly of concave cubes was reproducibly obtained by drop casting a droplet of nanoparticle solution containing AlCNCs in ambient conditions, offering a cost-effective and nanofabrication-free way to form a large area of plasmonic substrates [23, 24].

Machine learning has increasingly become a powerful tool in nanoscale science, surface chemistry, and biosensors, enabling advances in fabrication, characterization, and property prediction by integrating experimental data with physics-informed models. While ML has been applied to Raman and fluorescence spectroscopy for biochemical sensing, this paper is the first application of ML for the classification of neurotransmitters based on AFTDs. We performed comparative analysis using three ML techniques - Long Short-Term Memory (LSTM), K-Nearest Neighbors (KNN), and Random Forest (RF) [25-27]. LSTM on AFTDS collected on AlCNCs achieved the highest classification accuracy, followed by a slightly poorer performance by KNN and RF on AFTDS. KNN and RF on AF collected in the solution phase without plasmonic nanoparticles achieved poorer performances. Our results emphasize the importance of plasmonic-engineered AFTDS in achieving high classification accuracy among similarly structured MANTs. In addition, the superiority of LSTM over KNN and RF in analyzing time-dependent AF data was demonstrated.

**Material and Methods**

Aluminum concave nanocubes solution with nominal diameters of 80±9 nm was purchased from NanoComposix (particle concentration: $3.9 \times 10^{12}$ particles/mL, mass concentration: 2.8 mg/mL, with a surface area of 27.3 m²/g, and were dissolved in 1-propanol). Dopamine, 3,4-dihydroxyphenylacetic acid, and norepinephrine (>99.9%) were obtained from Sigma-Aldrich.

Fluorescence measurements of the neurotransmitters were conducted on two substrates: AlCNC and plain silicon wafers as a reference. The ALCNC structure was selected because the extinction values are relatively higher at the excitation/emission wavelengths of neurotransmitters compared to other commercially available aluminum nanoparticles. We also suspect that its concave geometry, featuring sharp corners and edges similar to those in aluminum bowtie nano antennas, nanotriangles, and hole-arrays [28, 29], enabling it to excite stronger fluorescence enhancement compared to other geometries without sharp corners or edges[21, 30]. A standard 2-inch silicon wafer was cut into four pieces and treated with a plasma cleaner for 90 seconds at a base pressure of 0.4 Torr to render the surface hydrophilic. Subsequently, 5 μL of the AlCNC nanoparticle (NP) solution was drop-cast onto the silicon substrate and allowed to dry in ambient conditions [23]; offering a simple and time-efficient fabrication route without requiring complex methods or equipment such as e-beam lithography or sputtering [21]. After drying, a multi-layer pattern of the nanoparticles remained on the substrate surfaces. The structural and compositional analysis of AlCNCs was performed using scanning electron microscopy (SEM) and scanning transmission electron microscopy (STEM). Following substrate preparation, solutions of three MANTs in deionized water at varying concentrations were prepared, and 1 μL of each solution was drop-cast [24] onto two different substrates. The outer edge of the coffee ring pattern of the AlCNC where high concentration of AlCNCs were found and a Si wafer to measure the fluorescence spectrum. The dried MANTs were then exposed to a 266 nm UV laser with an incident angle of 60 degrees, and AFTDS were collected using a Hitachi IHR 550 spectrometer coupled with a CCD camera [19-22]. The AFTDS were taken over 2 minutes and an exposure time of 0.5 seconds was used for each spectrum within the AFTDS.

We performed a comparative analysis using three ML techniques: First, we used LSTM; a type of recurrent neural network well-suited for sequence prediction, which can capture temporal dependencies in the fluorescence data, potentially improving the accuracy for distinguishing closely related biomolecules. While LSTM has been applied for capturing sequential patterns and forecasting time-series tasks such as molecular generation [31], property prediction [32, 33], and dynamic process modeling [34, 35], this paper presents the first application of LSTM in classifying molecules based on AFTDS. Secondly, we used KNN; a simple, instance-based learning algorithm that classifies data points based on the closest training examples in the feature space [36]. Third, we used RF, an ensemble learning method based on decision trees, to classify the spectral data and assess its performance compared to other algorithms [27]. The AFTDS data was recorded as a time-resolved spectrum spanning 280–360 nm, with each measurement capturing one spectrum every 0.5 seconds for 2 minutes, yielding 240 spectra per laser spot. These were converted into structured input tensors of shape (time_steps × wavelength_bins) for sequence-based models (LSTM) or flattened into feature vectors for non-sequential models (KNN, RF). Processing the AFTDS with ML was accomplished in different steps; the fluorescence data gathered from multiple experiments was used to train the ML model. Each input sample was labeled with the corresponding neurotransmitter class (DA, DOPAC, NE) to train supervised classifiers. The training process involved pre-processing the fluorescence data, feature extraction, and applying data augmentation techniques to improve the model's accuracy and generalizability.

Distinguishing neurotransmitters with similar structures solely based on their native fluorescence is challenging due to their overlapping fluorescence spectra. To address this challenge, we employed ML classification models and developed a reliable classification framework capable of differentiating neurotransmitters based on their unique fluorescence signatures. This approach allows for more precise detection and identification of neurotransmitters. The integration of ML not only improves classification accuracy but also provides a scalable method for analyzing large experimental datasets 6, 10.

We scaled all fluorescence data with MinMaxScaler and evaluated models with 4-fold cross-validation (75 % train, 25 % test) as Fig. 1 shows the workflow. For the in-solution emission spectra, each sample was a static snapshot of paired wavelength–intensity values fed directly to KNN and a 150-tree RF. For the AFTDS, we supplied KNN and RF with flattened, right-aligned intensity vectors, while the LSTM received the raw 17-step sequence, letting its three stacked LSTM layers learn temporal dependencies before batch normalization, a dense layer, and soft-max output. This unified preprocessing lets each model exploit either static spectral structure or time dynamics as appropriate, yielding consistent classification performance across modalities. Further details and the code used to process the data, fit the model, and generate predictions can be found in the data availability section.

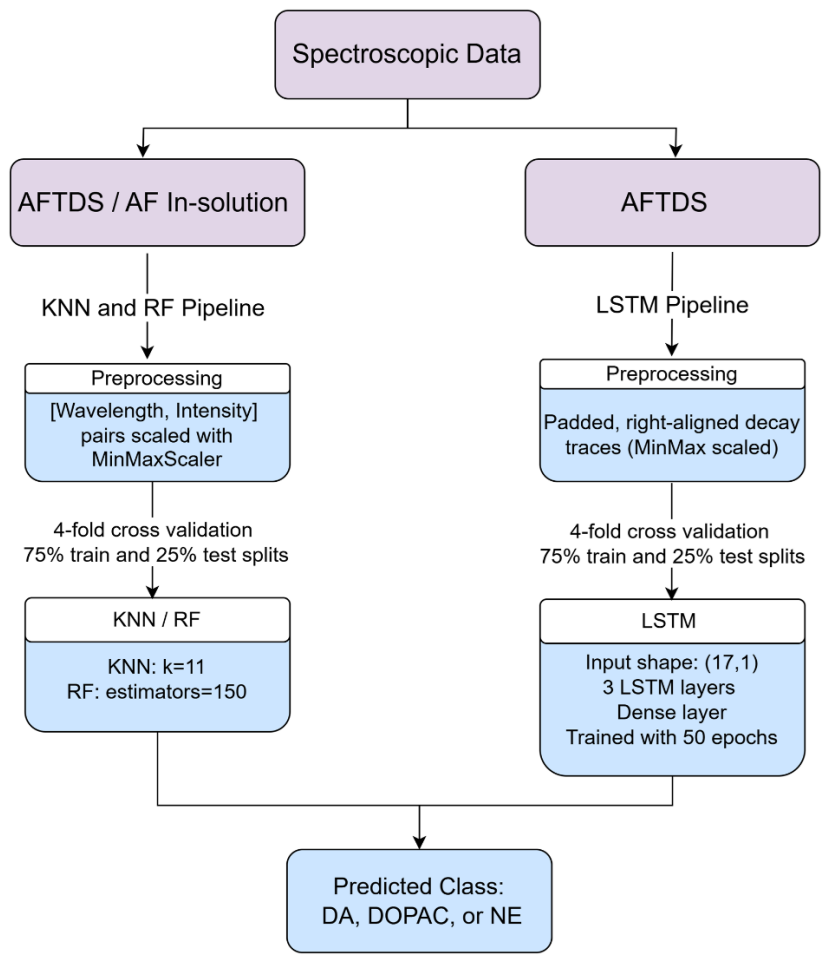

**Fig. 1.** Workflow for classifying DA, DOPAC, and NE from UV autofluorescence: MinMax-scaled snapshot spectra are routed to an 11-nearest-neighbour or 150-tree random-forest classifier, whereas 17-point AFTDS traces feed a three-layer LSTM; all models are assessed by 4-fold cross-validation and return the predicted neurotransmitter class.

The AFTDS collected on AlCNC substrates contain only seventeen regularly spaced intensity values per trace, so the data are sequential and short. A LSTM network suits this setting because its gated recurrent structure keeps information from both the fast initial quench and the slower tail of the decay curve, the two regions that together separate DA, DOPAC, and NE even though they occur at very different times.

The in-solution spectra, by contrast, are single wavelength-intensity snapshots. For these static measurements, we wanted algorithms that run quickly while still capturing subtle spectral patterns. An 11-neighbor KNN classifier provides a simple non-parametric baseline that classifies by proximity in wavelength-intensity space, while a RF with 150 trees captures non-linear interactions among neighboring wavelengths and offers built-in feature importance scores that aid interpretation.

Simpler alternatives such as logistic regression and single decision trees were rejected before testing. Linear decision boundaries cannot recover the class separation that appears only after part of the fluorescence has decayed, and a single tree is sensitive to noise in individual wavelengths. Either option would likely underfit or overfit without reducing inference time, since RF and KNN already respond well under a millisecond on the intended devices. Therefore, we concentrated our experiments on the LSTM model for the sequential traces and on KNN together with RF as fast, interpretable baselines for the static spectra.

Using TensorFlow 2.18.0 [30], we trained an LSTM model on AFTDS data for molecular classification, achieving 89% accuracy on both training and validation datasets. To ensure robust evaluation and address potential overfitting, 4-fold cross-validation was implemented. The training data was partitioned into four equal folds, with three folds (75%) used for training and one-fold (25%) reserved for validation in each iteration. This process was repeated four times, allowing each fold to serve as the validation set once. By exposing the model to diverse training and validation distributions, cross-validation effectively addressed class imbalance, enhanced generalization, and minimized bias.

Fig. 2 presents the training and validation curves of the LSTM model over 50 epochs, showing model accuracy (blue) and model loss (red). Both training and validation accuracy increase rapidly during the initial epochs and stabilize around 0.89, indicating good generalization and minimal overfitting. The k-fold cross-validation results further support the model's consistency, with accuracy ranging from 88.2% to 90.9% and loss values stabilizing around 0.3 across folds.

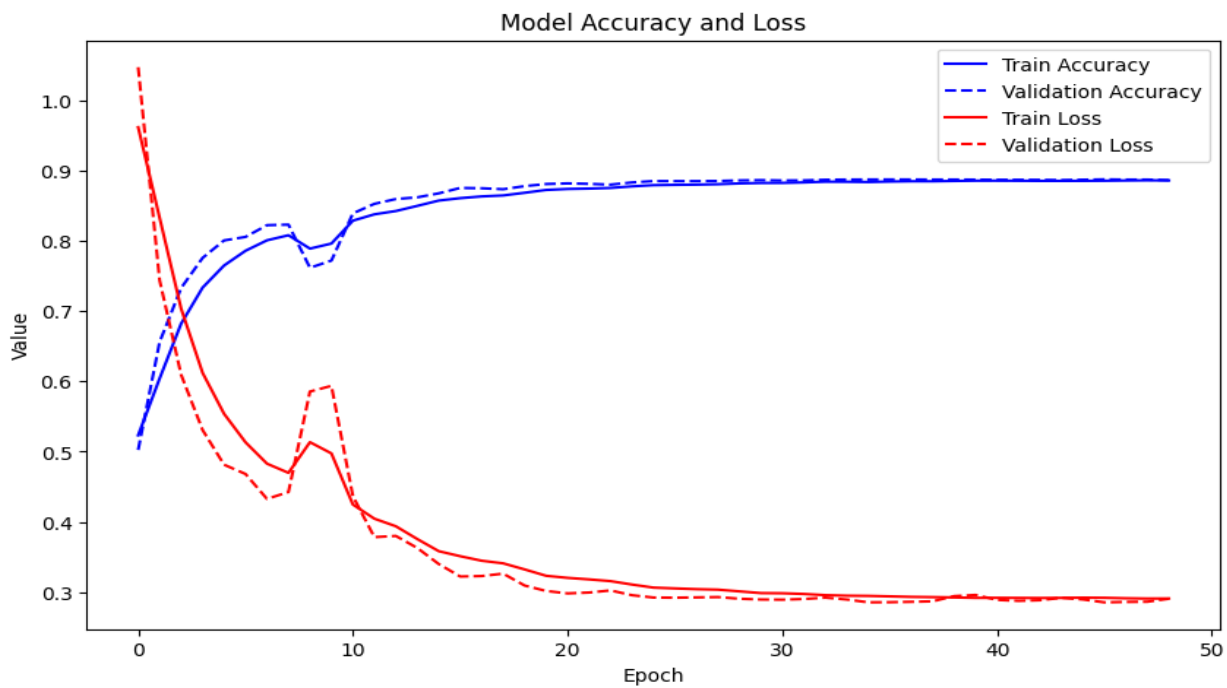

**Fig. 2.** The plot presents the training and validation curves of the LSTM model, illustrating model accuracy (blue) and model loss (red) over 50 training epochs.

The dataset used for model training and evaluation consisted of 146,432 samples for models trained on AlCNC and 4,113 samples for in-solution models. A 75%-25% split was applied, allocating approximately 109,824 samples for training and 36,608 samples for testing in the AlCNC dataset, while the in-solution dataset contained 2,817 training samples and 939 test samples the dataset captures both spectral and temporal fluorescence dynamics, providing a rich input for classification.

**Results and Discussion**

**Experimental Results**

Fig. 3 (A) shows a SEM image of the coffee ring pattern (the dark circle) left by a drop of deionized water containing molecules after it evaporated on the AlCNCs. The UV laser focuses near the dark circle where the concentration of molecules is higher than in other spots to maximize the fluorescence signal. Fig. 3 B shows the extinction spectrum of ALCNC measured by a UV-VIS spectrometer. The high extinction values at the excitation wavelength 266 nm and the emission peaks of neurotransmitters (300-320) nm are linked to strong plasmonic resonances at these wavelengths [37]. Fig. 3 (C-F) presents the STEM analysis of AlCNCs, highlighting their composition and surface features. The Energy Dispersive Spectroscopy (EDS) in Fig. 3C confirms aluminum (Al) as the dominant component, with oxygen (O) mainly on the surface of the AlCNCs. The average diameter of ALCNCs were estimated to be 80±9 nm by STEM tools, and the thickness of the oxide layer on the surface of AlCNCs after exposure to air is estimated to be between 4 to 8 nm (Fig. 3D). The cyan-colored EDS map represents aluminum in Fig. 3E, while the red map corresponds to the oxide layer (Fig. 3F). These results provide a detailed characterization of the AlCNCs and their surface properties. SEM images of ALCNC nanoparticles, along with the coffee ring pattern formed by drop-casting 1 µL of molecule solution, are also presented in Figure S1.

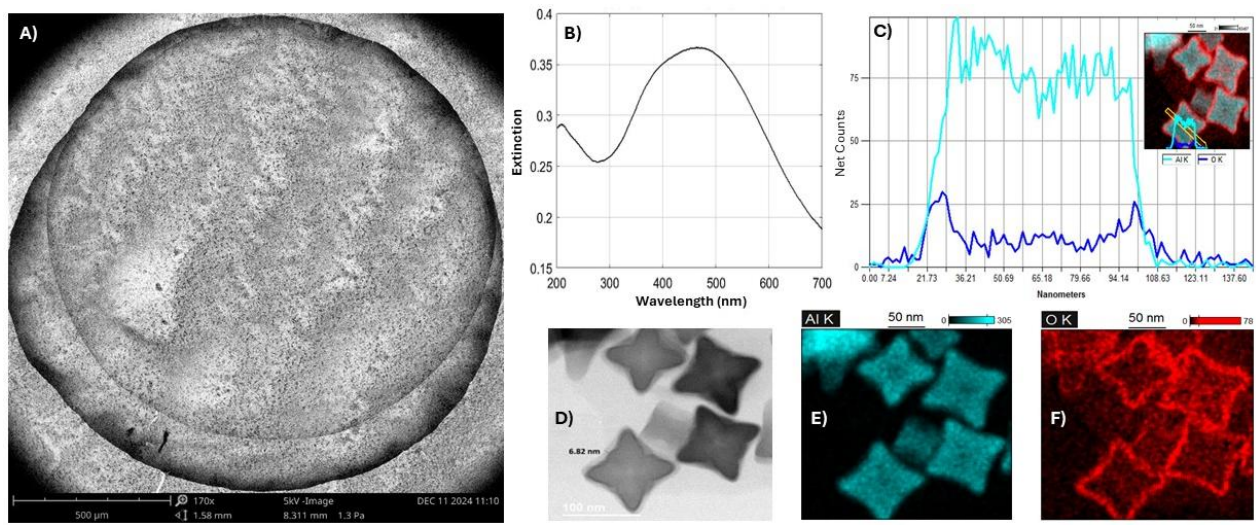

**Fig 3.** The characterization of aluminum concave nanocubes. (A) A coffee ring formed by evaporating a drop of deionized water containing molecules on ALCNCs. (B) Extinction spectrum of AlCNC (C) EDS line scan profile over a selected region of an AlCNC, illustrating the distribution of Al (cyan) and O (blue), with the inset displaying the scan area superimposed on an AlCNC. (D) STEM images highlight the morphology and concave geometry of the AlCNCs, with oxide layer thickness estimated to be 6±2 nm. (E-F) EDS elemental mapping images showing the spatial distribution of Al (cyan) and the oxide layer (red) on the AlCNC surfaces.

The extinction coefficients ($\varepsilon$) of DA, NE, and DOPAC were determined from the slope of linear fit data of absorption vs. concentration plots in Fig. 4, using five Quantum Yield (QY: the percentage of photons absorbed that convert to photons emitted) data points in our previous work. The calculated $\varepsilon$ were 2110, 1070, and 2210 Lmol$^{-1}$cm$^{-1}$ for DA, NE, and DOPAC, respectively, as presented in Table 1. Table 1 also lists the QY values of these molecules from our previous publication [20]. The fluorescent data for DA, NE, and DOPAC in deionized water at 5 different concentrations are shown in Figure S2, and the corresponding absorption data are shown in Figure S3.

The fluorescence intensity is a product of QY and absorption coefficients (proportional to the number of photons absorbed). The absorption coefficients are usually estimated by $\varepsilon$ (absorption plus scattering). The lower fluorescence for DOPAC is due to the lower QY of DOPAC compared to DA and NE. While a molecule with a higher QY generally exhibits greater fluorescence intensity [20], it is important to also consider the product of QY and $\varepsilon$ (QY × $\varepsilon$). The values of QY × $\varepsilon$ for the DA, NE and DOPAC in deionized water are listed in Table 1. For example, although NE has a slightly higher QY than DA (6.3% vs. 5.9%), DA exhibits a significantly higher QY × $\varepsilon$ value (124.5 vs. 67.4), which corresponds with its stronger fluorescence signal observed in our measurements. In contrast, although DOPAC exhibits an $\varepsilon$ comparable to DA, its much lower QY (1.7%) leads to significantly reduced fluorescence intensity [38, 39].

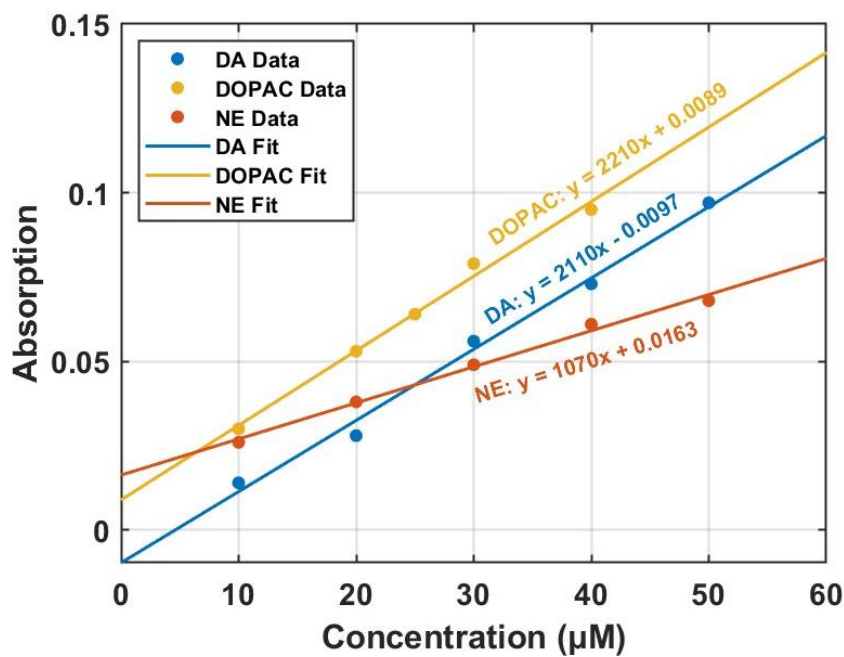

**Fig 4.** Absorption vs. concentration data fit for DA, DOPAC, and NE in deionized water.

Table 1 - Quantum Yield and Extinction Coefficients of DA, DOPAC, and NE

| Molecule | QY, Quantum Yield (%) | ε, Extinction Coefficient (L/mol.cm) | ε × QY |
|---|---|---|---|
| **DA** | 5.9 ± 0.8 [20] | 2110 ± 111.2 | 124.5 ± 18.1 |
| **NE** | 6.3 ± 0.6 [20] | 1070 ± 117.6 | 67.4 ± 9.8 |
| **DOPAC** | 1.7 ± 0.2 [20] | 2210 ± 52.6 | 37.6 ± 4.5 |

We employed AlCNCs that were drop cast and dried on a bare Si wafer (as reference) as plasmonic substrates to differentiate and analyze three similar neurotransmitters: DA, DOPAC, and NE. Fig. 5 (A, B, C) presents AFTDS signals, decreasing over time for DA, DOPAC, and NE, and the topmost curve (highest intensity) corresponds to the first spectrum. The AF spectra, shown in Fig. 5D and Fig. 5E, compare the intensities of DA, DOPAC, and NE acquired at 0–0.5 seconds with an acquisition time of 0.5 seconds ($I_0$, the highest intensity spectrum in the AFTDS) on a silicon wafer (Fig. 5D) and AlCNC substrates (Fig. 5E). DA exhibited the highest fluorescence intensity on both substrates, followed by NE and DOPAC. Notably, fluorescence intensities on the AlCNC substrate were significantly higher compared to a silicon substrate, highlighting the fluorescence-enhancing properties of AlCNCs. This enhancement is specifically attributed to the localized surface plasmon resonance (LSPR) effects of AlCNCs in the UV range, which amplifies both the excitation and fluorescent emission near the nanocube surface, unlike the non-plasmonic silicon substrate [38]. Additionally, fluorescence spectra data acquired using a fluorometer for the neurotransmitters dissolved in water are presented in Fig. 5F.

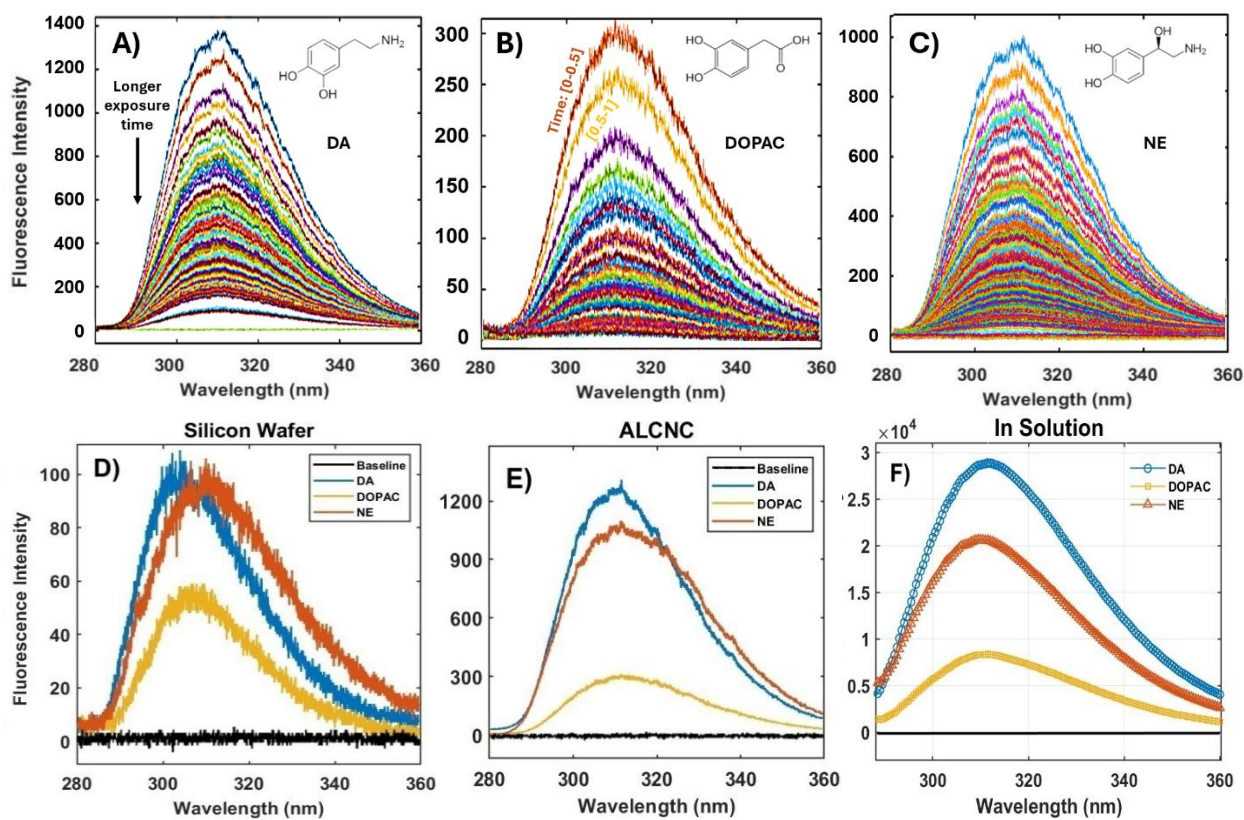

**Fig. 5.** AFTDS for three neurotransmitters A) DA, B) DOPAC, and C) NE drop cast and dried on AlCNC substrates (1µL of 500 µM). Fluorescence intensity of DA, DOPAC, and NE (1µL of 500 µM) collected at 0 to 0.5 seconds on D) a silicon wafer and E) ALCNC substrates, F) Fluorescence Spectra of DA, DOPAC, and NE solution dissolved in water at a concentration of 50 µM measured by a fluorometer.

Fig. 6 compares the average integrated fluorescence intensities and net enhancement factors for DA, NE, and DOPAC; each droplet contains 1µL of 500 µM of MANTS that drop cast and dried on a silicon wafer and AlCNC substrates. The integrated fluorescence intensity and net enhancement factors were calculated using the methods published previously [19, 20]. To quantify the variability in enhancement measurements, the error was calculated as the standard deviation (STD) across multiple experimental replicates [39]. For each neurotransmitter (DA, DOPAC, and NE), fluorescence spectra were collected at multiple distinct spots (n = 5–7) on both AlCNC and silicon (Si) substrates. The integrated fluorescence intensity for each measurement was obtained by subtracting the baseline signal (dark spectrum) and computing the area under the spectrum using the trapezoidal rule. Enhancement factors were calculated by normalizing the integrated intensity on AlCNC to the corresponding average intensity on Si. The reported enhancement values represent the mean of these normalized measurements, and the associated errors reflect the standard deviation across the set of replicates, thereby accounting for experimental variation due to sample heterogeneity, measurement noise, and substrate uniformity. As shown in Fig. 6A, fluorescence intensities on the AlCNC substrate were significantly higher than those on the silicon wafer, with DA exhibiting the strongest signal, followed by NE and DOPAC. Fig. 6B highlights the net enhancement factors achieved on AlCNC, with DA, NE, and DOPAC showing enhancements of 12, 9, and 7 respectively, compared to silicon wafers.

These results demonstrate that AlCNCs drop cast and dried on a substrate are effective in enhancing the AF signal of MANTs. Although other plasmonic nanostructures have achieved higher fluorescence enhancement factors, the presented method achieves a plasmonic substrate without complicated, lengthy and expensive nanofabrication techniques.

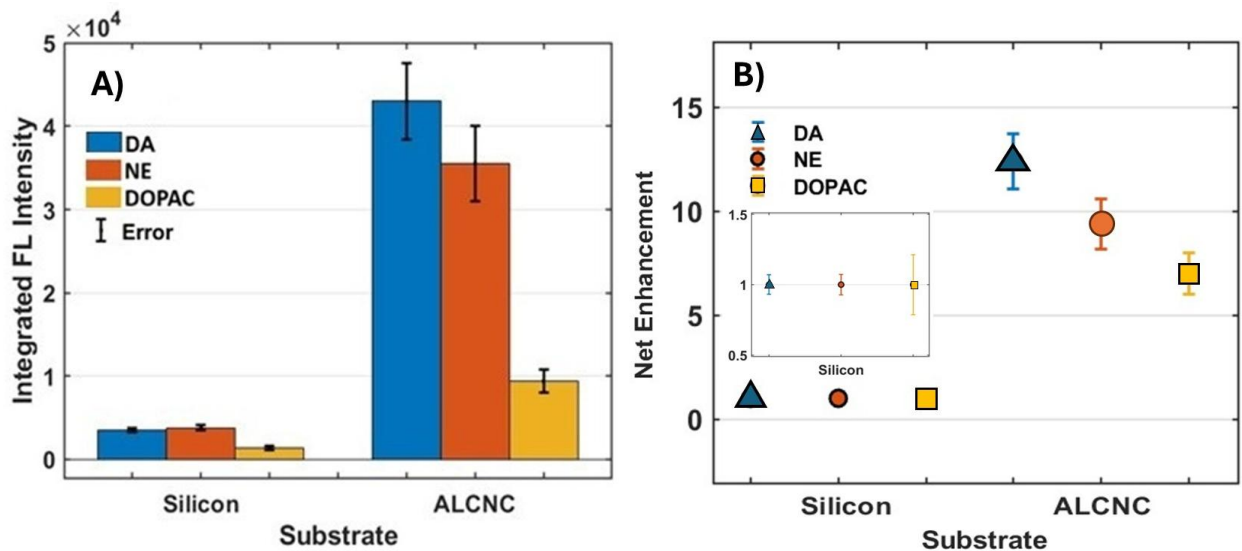

**Fig. 6.** A) Integrated Fluorescence Intensity and B) Net Enhancement factors from three neurotransmitters: DA, DOPAC, and NE, comparing signal levels between AF data on silicon wafers and AlCNCs.

The AFTDS data on different concentrations for model training and evaluation are shown in Figure S4. The dataset contained three classes—DA, DOPAC, and NE—with slightly varying support: 54,784, 51,712, and 39,936 samples, respectively, in the AlCNC dataset, and 1105, 1,326, and 1,326 samples in the in-solution dataset and more details about samples in each class can be found in Figure S5. Each sample corresponds to a unique fluorescence data point defined as a pair of fluorescence intensity and wavelength values, represented as a function FL(I, Wv). As shown in Figure S4, each fluorescence time series was acquired over about 2-minute measurement timeframe with a 0.5-second acquisition interval, yielding approximately 240 spectra per measurement. Each of these spectra spans a 60 nm range (280–360 nm), representing the AFTDS for a specific molecule at a defined concentration. The weighted and macro averages in performance metrics were used to ensure a balanced evaluation across all classes.

**Classification Results**

Fig. 7 presents the confusion matrices for the LSTM, KNN, and RF models used to classify three neurotransmitters—DA, DOPAC, and NE—based on AF data obtained from AlCNC and in-solution environments. The matrices display class-specific prediction distributions, with diagonal elements representing correct or true positives and off-diagonal elements indicating misclassifications. The LSTM model on AFTDS achieved high classification accuracy, with 88.0% for DA, 90.4% for DOPAC, and 89.2% for NE, indicating strong temporal feature learning. The KNN model on AFTDS also demonstrated high performance, achieving 89.3% for DA, 86.1% for DOPAC, and 83.0% for NE. Similarly, the RF on AFTDS achieved 87.4% for DA, 85.6% for DOPAC, and 76.7% for NE, supporting its strong ensemble classification ability on temporal AFTDS data. Although RF outperformed KNN on the AF in-solution dataset (RF achieving 60.9% for DA, 82.1% for DOPAC, and 68.6% for NE versus KNN's 44.8% for DA, 89.0% for DOPAC, and 66.6% for NE), both models performed substantially worse compared to those applied to AFTDS data. This drop in accuracy highlights increased confusion among class boundaries in the solution-based measurements. The absolute number of samples per matrix cell was provided inside the brackets in Figure S5 and in Table S1, revealing the distribution and support for each class. These matrices provide a clear visual assessment of model effectiveness and misclassification tendencies.

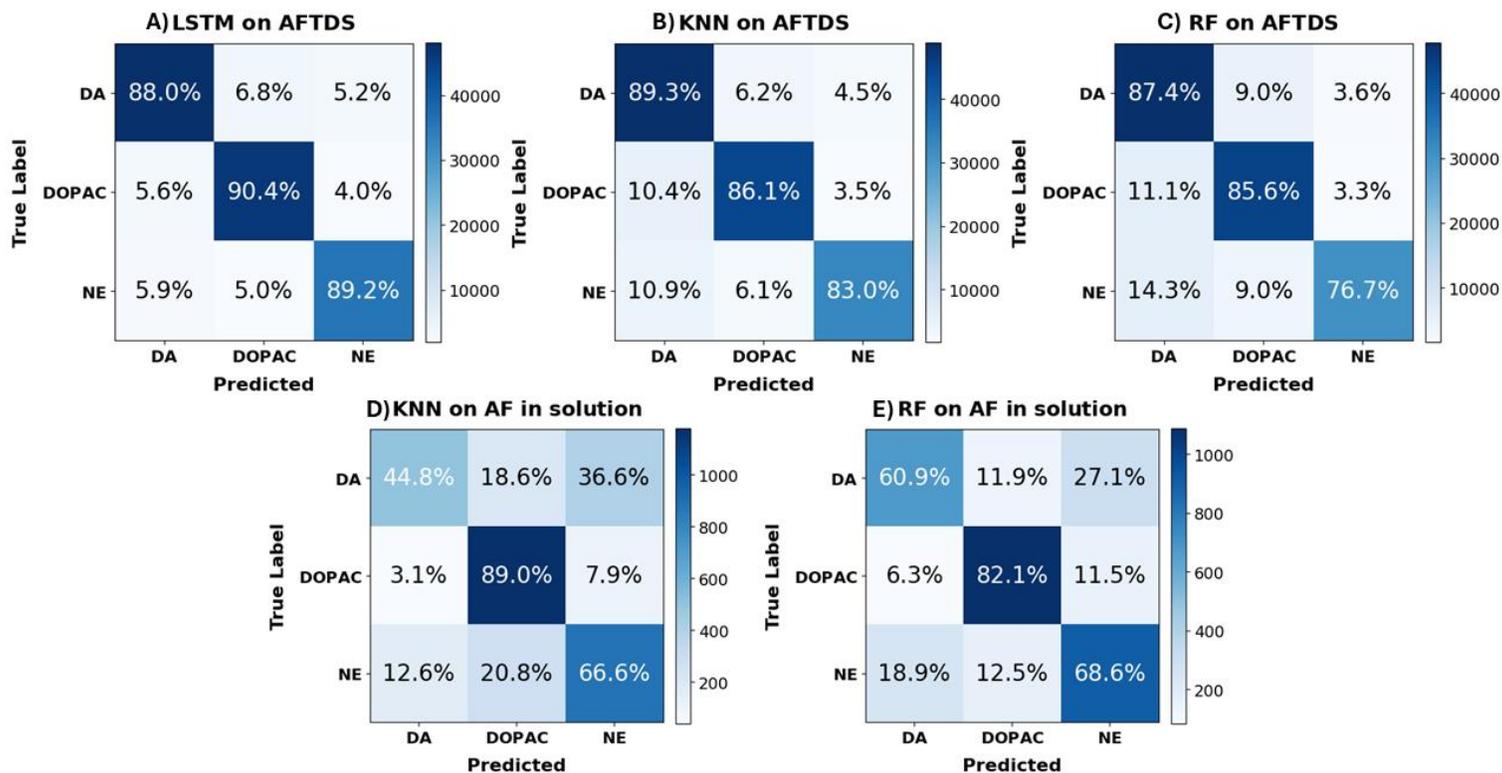

**Fig. 7.** Confusion matrices for LSTM, KNN and RF classifiers distinguishing DA, DOPAC and NE from AFTDS and AF in-solution data. Panels A–C (AFTDS) retain high diagonal accuracies, whereas the accuracies in panels D–E (AF in-solution) drop sharply, especially for DA and NE, signaling greater inter-class confusion. Cell values are percentages; diagonals denote correct predictions, and off-diagonals indicate misclassifications.

The classification performance metrics, Precision, Recall, and F1 Score, for LSTM, KNN, and RF models across three molecular classes: DA, DOPAC, and NE, are presented in Table 2. These metrics, derived from confusion matrices, provide insight into each model's classification, accuracy, and efficiency. We included **macro** and **weighted averages** to summarize model performance across all classes. **Macro average** treats all classes equally, useful for balanced datasets, while **the weighted average reflects class distribution, favoring the** majority classes. These metrics offer complementary views but can mislead if used alone in imbalanced settings. For better clarity, bar charts are provided in Figure S6.

Table 2. Classification Models Performance Metrics Across Classes (DA, DOPAC, NE). The best-performing values for each metric and class within each dataset are **bolded and underlined** for emphasis. Values for recall are also shown in the confusion matrices.

| Model | Class | DA | NE | DOPAC | Weighted Avg | Macro Avg | Accuracy |
|---|---|---|---|---|---|---|---|
| LSTM | Precision | **0.90** | 0.88 | **0.89** | **0.89** | 0.89 | - |
|  | Recall | 0.88 | **0.89** | 0.90 | **0.89** | **0.89** | - |
| using AFTDS of MANTs on AlCNC | F1 Score | **0.89** | **0.88** | **0.90** | **0.89** | **0.89** | **0.89** |
| KNN using AFTDS of MANTs on AlCNC | Precision | 0.83 | 0.88 | 0.89 | 0.87 | 0.87 | - |
| | Recall | **0.89** | 0.83 | 0.86 | 0.87 | 0.86 | - |
| | F1 Score | 0.86 | 0.86 | 0.87 | 0.87 | 0.87 | 0.86 |
| RF using AFTDS of MANTs on AlCNC | Precision | 0.81 | **0.89** | 0.84 | 0.84 | 0.85 | - |
| | Recall | 0.87 | 0.77 | 0.86 | 0.84 | 0.83 | - |
| | F1 Score | 0.84 | 0.82 | 0.85 | 0.84 | 0.84 | 0.84 |
| KNN using AF of MANTs in the solution | Precision | 0.70 | 0.63 | 0.71 | 0.68 | 0.68 | - |
| | Recall | 0.45 | 0.67 | 0.89 | 0.67 | 0.68 | - |
| | F1 Score | 0.55 | 0.65 | 0.79 | 0.66 | 0.67 | 0.68 |
| RF using AF of MANTs in the solution | Precision | 0.67 | 0.67 | 0.78 | 0.71 | 0.71 | - |
| | Recall | 0.61 | 0.69 | 0.82 | 0.71 | 0.71 | - |
| | F1 Score | 0.64 | 0.68 | 0.80 | 0.71 | 0.71 | 0.71 |

The results demonstrate that the LSTM model consistently outperforms the KNN model when applied to AFTDS data collected from MANTs on AlCNCs. The LSTM model achieves a weighted and macro average F1 score of 0.89, indicating strong generalization and superior classification performance across all classes. In contrast, the KNN model yield a slightly lower weighted and macro average F1 score of 0.86, reflecting its comparatively reduced ability to capture dynamic spectral features.

For AF spectra data collected from MANTs in solution (without plasmonic nanostructures), the KNN model exhibits reduced classification performance compared to its performance on AFTDS data. It achieves a weighted and macro average F1 score of 0.68. Notably, KNN performs best on DOPAC but struggles with NE and exhibits the weakest performance on DA. This discrepancy may be attributed to differences in molecular intensity within solution-based measurements. Conversely, classification accuracy using AFTDS data remains relatively consistent across different molecules, underscoring its robustness in molecular classification.

In addition to KNN and LSTM, a RF model was evaluated under the same conditions. On AFTDS data, the RF model achieved a weighted and macro average F1 score of 0.84, slightly lower than the KNN and LSTM models but still demonstrating strong performance. When applied to AF data collected in solution, the RF model outperformed KNN, with a weighted and macro average F1 score of 0.71. This indicates that while RF does not match LSTM performance on AFTDS data, it provides more reliable classification than KNN in solution-based environments, likely due to its ensemble-based structure and resilience to data noise.

The sample size for data collected on AlCNC is much larger than that of solution-based spectra. Since ML benefits from large data sets, our methods of collecting time variant spectra under continuous light illumination are advantageous compared with solution-based spectra collection. For instance, using AlCNC and acquiring AFTDS data with an acquisition time of 0.5 seconds, we obtained 240 spectra from a single molecular concentration in 2 minutes. In contrast, obtaining the same number of spectra with solution-based UV-VIS would require 240 different concentrations and separate measurements, which would be an impractical, time-consuming, and costly approach.

Overall, these findings highlight the LSTM model's advantage in leveraging the time-dependent characteristics of AFTDS signals, allowing for more accurate molecular differentiation. In contrast, the KNN model, which relies on static data, demonstrates lower classification performance. The confusion matrix results further reinforce the LSTM model's effectiveness in distinguishing structurally similar MANTs based on AFTDS data while also showing the RF model as a better alternative than KNN when temporal data is limited or unavailable.

**Conclusion**

This work demonstrates a transformative approach to probe-free and label-free biosensing by merging autofluorescence using novel plasmonic nanoparticles with ML. Aluminum concave nanocubes, a UV plasmonic substrate, significantly enhanced the native fluorescence signals of neurotransmitters, achieving up to a 12-fold improvement compared to silicon substrates. This enhancement amplified the autofluorescence intensity signals, and AFTDS was the key factor that enabled the differentiation of similar neurotransmitter data in a label-free manner. When analyzed using ML techniques, the AFTDS on AlCNCs provided high classification accuracy, highlighting the critical role of plasmonic-engineered fluorescence dynamics in molecular identification.

Integrating ML models, particularly LSTM networks, proved critical in analyzing and classifying complex fluorescence data. With a classification accuracy of 89%, the ML models effectively captured subtle variations in spectral data, enabling the reliable identification of neurotransmitters. The study also highlighted the comparative advantage of LSTM over KNN and RF in handling time-dependent fluorescence data (AFTDS). In addition, RF models demonstrated competitive performance, particularly in environments where temporal patterns are less pronounced. While not as effective as LSTM on AFTDS data, the RF model surpassed KNN in classifying solution-based spectra, emphasizing its robustness and versatility as a static-data classifier and suggesting RF as a better approach in scenarios where dynamic signal acquisition may be limited.

These findings underscore the potential of combining nanotechnology and ML to create smart biosensing systems that are both sensitive and selective. The proposed methodology paves the way for developing non-invasive, real-time diagnostic tools for complex biological environments, offering new avenues for applications in biomedical diagnostics, neuroscience, and beyond.

**Author Contributions**

**Mohammad Mohammadi:** Conceptualization, Data curation, Formal analysis, Investigation, Methodology, Software, Validation, Visualization, Writing – original draft. **Sima Najafzadehkhoei:** Data curation, Software, Visualization, Methodology, Writing – original draft.

**George Vega Yon:** Writing – review & editing. **Yunshan Wang:** Supervision, Project administration, Resources, Investigation, Conceptualization, Methodology, Funding acquisition, Formal analysis, Writing – review & editing.

**Data Availability**

The data supporting this article have been included as part of the Supplementary Information. Also, the ML codes, preprocessed datasets, and training scripts used in this study have been made publicly available via [GitHub](https://github.com/sima-njf/Machine-Learning-on-UV-plasmonic-engineered-Auto-Fluorescence-Time-Decay-Series-AFTDS-/tree/main) at [https://github.com/sima-njf/Machine-Learning-on-UV-plasmonic-engineered-Auto-Fluorescence-Time-Decay-Series-AFTDS-/tree/main] in compliance with open access and reproducibility standards in ML-based research.

**Conflicts of Interest**

There are no conflicts to declare.

**Acknowledgment**

Funding for this project was provided by the University of Utah Technology Licensing Office (TLO) through the Ascender Grant. This work was performed in part at the Utah Nanofab Cleanroom and made use of Nanofab EMSAL shared facilities of the Micron Technology Foundation Inc. Microscopy Suite sponsored by the John and Marcia Price College of Engineering College of Engineering and the Office of the Vice President for Research. The author(s) appreciate the support of the staff and facilities that made this work possible.

**References**

1. R. I. Teleanu, A.-G. Niculescu, E. Roza, O. Vladâcenco, A. M. Grumezescu and D. M. Teleanu, *International journal of molecular sciences*, 2022, **23**, 5954.
2. H. A. Asrari and M. Peters, 2024.
3. F. Shojaeianforoud and M. Lahooti, *Computers in Biology and Medicine*, 2025, **190**, 110047. https://doi.org/10.1016/j.compbiomed.2025.110047
4. F. Shojaeianforoud, L. Marin, W. J. Anderl, M. Marino, B. Coats and K. L. Monson, *Acta Biomaterialia*, 2025, **197**, 256-265. https://doi.org/10.1016/j.actbio.2025.03.027
5. S. Greco, W. Danysz, A. Zivkovic, R. Gross and H. Stark, *Analytica chimica acta*, 2013, **771**, 65-72.
6. S. D. Niyonambaza, P. Kumar, P. Xing, J. Mathault, P. De Koninck, E. Boisselier, M. Boukadoum and A. Miled, *Applied sciences*, 2019, **9**, 4719.
7. G. P. Lachance, D. Gauvreau, É. Boisselier, M. Boukadoum and A. Miled, *Sensors*, 2024, **24**.
8. M. Shin and B. J. Venton, *Angewandte Chemie*, 2022, **134**, e202207399.
9. C. Zhang, X. You, Y. Li, Y. Zuo, W. Wang, D. Li, S. Huang, H. Hu, F. Yuan and F. Shao, *Sensors and Actuators B: Chemical*, 2022, **354**, 131233.

10. F. Sun, J. Zeng, M. Jing, J. Zhou, J. Feng, S. F. Owen, Y. Luo, F. Li, H. Wang and T. Yamaguchi, *Cell*, 2018, **174**, 481-496. e419.
11. A. G. Salinas, J. O. Lee, S. M. Augustin, S. Zhang, T. Patriarchi, L. Tian, M. Morales, Y. Mateo and D. M. Lovinger, *Nature communications*, 2023, **14**, 5915.
12. X. Fan, Y. Feng, Y. Su, L. Zhang and Y. Lv, *RSC Advances*, 2015, **5**, 55158-55164.
13. W. Hu, Y. Huang, C. Chen, Y. Liu, T. Guo and B.-O. Guan, *Sensors and Actuators B: Chemical*, 2018, **264**, 440-447.
14. C. Zhao, T. Man, Y. Cao, P. S. Weiss, H. G. Monbouquette and A. M. Andrews, *ACS sensors*, 2022, **7**, 3644-3653.
15. X. Liu, F. He, F. Zhang, Z. Zhang, Z. Huang and J. Liu, *Analytical chemistry*, 2020, **92**, 9370-9378.
16. K. M. Cheung, K.-A. Yang, N. Nakatsuka, C. Zhao, M. Ye, M. E. Jung, H. Yang, P. S. Weiss, M. N. Stojanovic and A. M. Andrews, *ACS sensors*, 2019, **4**, 3308-3317.
17. N. Nakatsuka, J. M. Abendroth, K.-A. Yang and A. M. Andrews, *ACS applied materials & interfaces*, 2021, **13**, 9425-9435.
18. Y. Zhang, W. Ren, Y. Z. Fan, H. Q. Luo and N. B. Li, *Sensors and Actuators B: Chemical*, 2020, **305**, 127463.
19. J.-Y. Lee, X. Cheng and Y. Wang, *Journal of Physics D: Applied Physics*, 2021, **54**, 135107.
20. J.-Y. Lee, M. Mohammadi and Y. Wang, *RSC advances*, 2023, **13**, 32582-32588.
21. M. Mohammadi and Y. Wang, 2024.
22. M. Mohammadi, B. Zhao, S. Salehi, O. Kingstedt, Y. Wang and S. Pan, *Chemical Communications*, 2024, **60**, 2780-2783.
23. M. Martinez-Garcia, P. Cardoso-Avila and J. Pichardo-Molina, *Colloids and Surfaces A: Physicochemical and Engineering Aspects*, 2016, **493**, 66-73.
24. M. M. Martinez-Garcia, J. L. Pichardo-Molina, N. Arzate-Plata and J. J. Alvarado-Gil, *Nanomaterials and Nanotechnology*, 2022, **12**, 18479804221082778.
25. I. Malashin, V. Tynchenko, A. Gantimurov, V. Nelyub and A. Borodulin, *Polymers*, 2024, **16**, 2607.
26. H. Flores-Hernandez and E. Martinez-Ledesma, *Journal of Cheminformatics*, 2024, **16**, 129.
27. V. Svetnik, A. Liaw, C. Tong, J. C. Culberson, R. P. Sheridan and B. P. Feuston, *Journal of chemical information and computer sciences*, 2003, **43**, 1947-1958.
28. X. Cheng, M. Rodriguez and Y. Wang, *Physica Scripta*, 2023, **98**, 115911.
29. X. Cheng, E. Lotubai, M. Rodriguez and Y. Wang, *OSA Continuum*, 2020, **3**, 3300-3313.
30. X. Cheng, M. Rodriguez and Y. Wang, 2020.
31. X. Xu, W. Huo, F. Li and H. Zhou, *Sensors*, 2023, **23**, 1149.
32. L. Xu, S. Pan, L. Xia and Z. Li, *Biomolecules*, 2023, **13**, 503.
33. M. Motamedi, R. Shidpour and M. Ezoji, *Scientific Reports*, 2024, **14**, 24353.
34. R. Kausar, F. Zayer, J. Viegas and J. Dias, 2024.
35. B. Xu, Y. Wang, L. Yuan and C. Xu, *Applied Intelligence*, 2023, **53**, 1619-1639.
36. C. Hulett, A. Hall and G. Qu, 2012.
37. S. A. Maier, *Plasmonics: fundamentals and applications*, Springer, 2007.
38. Y. Wang, E. M. Peterson, J. M. Harris, K. Appusamy, S. Guruswamy and S. Blair, *The Journal of Physical Chemistry C*, 2017, **121**, 11650-11657.
39. J. R. Lakowicz, *Principles of fluorescence spectroscopy*, Springer, 2006.